\documentclass[fleqn,usenatbib]{mnras}
\usepackage{newtxtext}
\usepackage{mathptmx}
\usepackage{mathtools}
\usepackage{txfonts}
\usepackage[T1]{fontenc}
\usepackage{ae,aecompl}
\usepackage{xcolor}

\DeclareRobustCommand{\VAN}[3]{#2}
\let\VANthebibliography\thebibliography
\def\thebibliography{\DeclareRobustCommand{\VAN}[3]{##3}\VANthebibliography}

\usepackage{graphicx}	
\usepackage{amsmath}	
\usepackage{amssymb}
\usepackage{bm}
\usepackage{pdflscape}

\title{Erratum: Primordial black hole origin for thermal gamma-ray bursts}

\author[O. del Barco]{Oscar del Barco\thanks{E-mail: obn@um.es}
\\
Laboratorio de \'{O}ptica, Instituto Universitario de Investigaci\'{o}n en \'{O}ptica y Nanof\'{i}sica, Universidad de Murcia, Campus de Espinardo, E-30100, Murcia, Spain\\}

\pubyear{2022}

\begin{document}
\label{firstpage}
\pagerange{\pageref{firstpage}--\pageref{lastpage}}
\maketitle


The paper \textit{Primordial black hole origin for thermal gamma ray bursts}
was published in MNRAS, 506, 806–812 (2021). In the original article,
an error in equation~(14) was found. Once taken into account that
the ratio $R_{\nu}/\nu^{3}$ is Lorentz-invariant \citep{Misner1973},
the correct expression for the observed spectral radiance should be rewritten as
\begin{equation}\label{Rnu}
R_{\nu} = \frac{2\pi h \nu^{3}}{c^{2}}
\left(\exp\left[\frac{h\nu}{k T_{\rm P}^{({\rm G})}}\right]
-1\right)^{-1},
\end{equation}
where $T_{\rm P}^{({\rm G})}$ is given by equation~(16)
(with an appropriate modification of
the emission angle, as described later). Accordingly,
equations~(15) and (19) should be modified by the following expressions
\begin{equation}\label{Wieng}
\nu_{\rm max} = \left(\frac{2.821}{h}\right)
k T_{\rm P}^{({\rm G})}.
\end{equation}
and
\begin{equation}\label{nuSnu}
\nu S_{\nu} = \frac{2\pi h \Omega \nu^{4}}{c^{2}}
\left(\exp\left[\frac{h\nu}{k T_{\rm P}^{({\rm G})}}\right]-1\right)^{-1}.
\end{equation}
As a consequence, the original selection of the initial
parameters resulted in an improbable observational scenario.
Nevertheless, the fundamental theory and
conclusions of the original article remain the same.

\begin{figure*}
\begin{center}
\includegraphics[width=.71\textwidth]{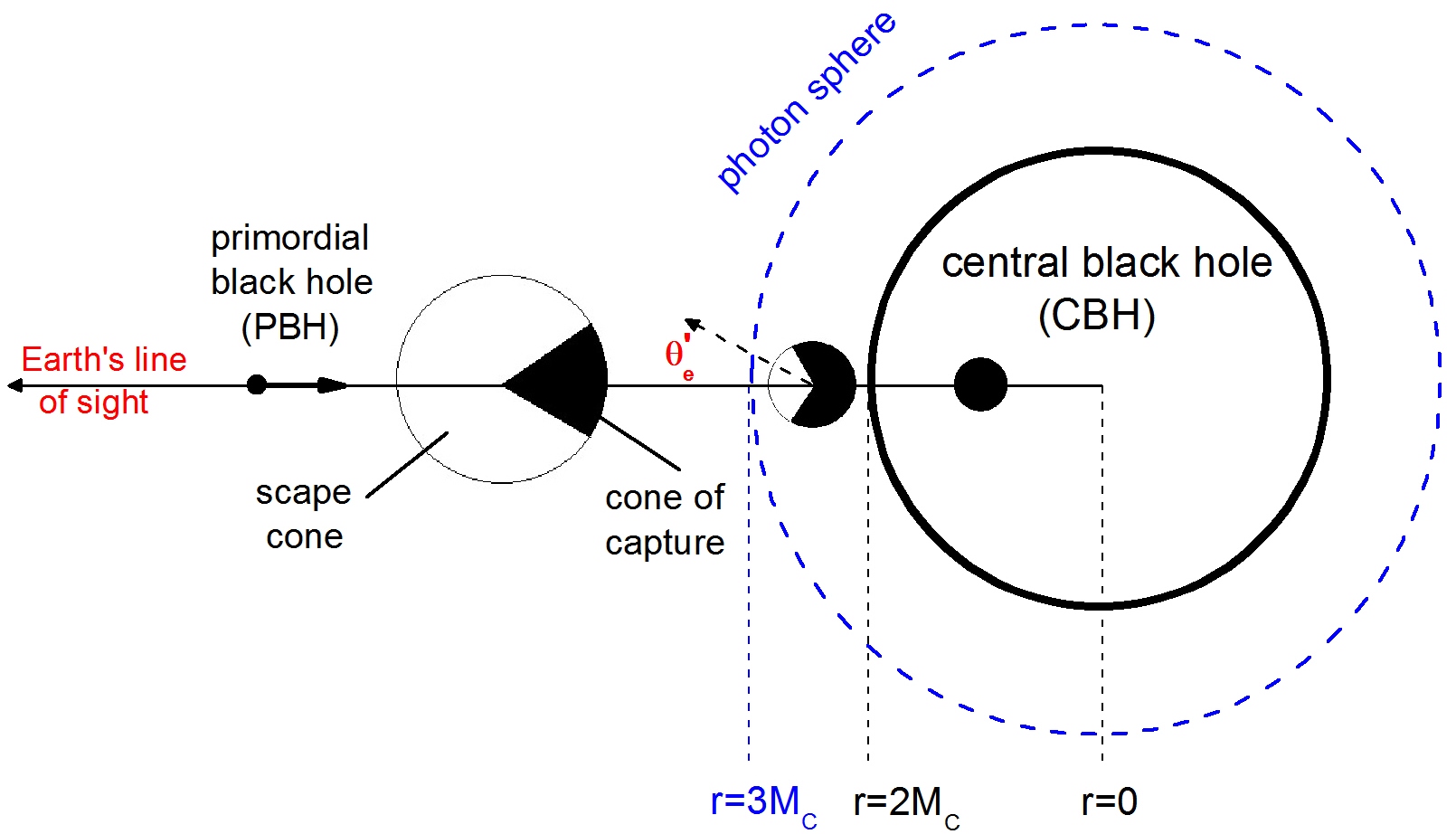}
\caption{Schematic representation of our
astrophysical scenario. The emission angle as observed from the PBH frame
$\theta_{\rm e}^{'}$ is shown. No critical angle $\theta_{\rm c}^{'}$ is
considered in the model.}
\label{fig1}
\end{center}
\end{figure*}

In Fig.~\ref{fig1}, the emission angle as observed from the PBH frame
$\theta_{\rm e}^{'}$ is shown. Compared to previous set-up,
no critical angle $\theta_{\rm c}^{'}$ is
considered in the model, since photons emitted at that angle will
rarely reach the Earth. Moveover, under realistic initial parameters
(described next), the measurable PBH Hawking radiation will be restricted
to a region above the CBH photon sphere.

In relation to the PBH Hawking temperature for a distant observer
$T_{\rm P}^{({\rm G})}$, a recent publication by \cite{McMaken2022}
alleged a supposed inconsistency with equation~(16) in the original paper
\citep{Barco2021}. Let us show that there is no discrepancy.

Following the formalism by \cite{Yoshino2019} for a star
under gravitational collapse (where the star emission is modelled
as a blackbody emitter, receding away from the Earth),
the redshift factor $\alpha$ which relates the Planck
temperature as measured by the emitter and observer
(i.e., $T_{\rm P}^{({\rm G})} = \alpha T_{\rm P}^{'}$ in our case)
can be expressed as (equation~(49) of this reference)
\begin{equation}\label{red1}
\alpha = \sqrt{\frac{f(R)}{f(r_{\rm o})}} \left(1-\cos\theta_{\rm e}^{'}
\sqrt{1-\frac{f(r_{\rm e})}{f(R)}}\right),
\end{equation}
where $r_{\rm e}$ ($r_{\rm o}$) are the coordinate positions for the
emitter (observer), $M$ and $R$ correspond to
the mass and radius of the static star,
respectively, and $f(r)= 1 - (2M/r)$.

Taking into account equation~(39) of \cite{Yoshino2019}
for the velocity $\beta$ of the star surface measured in the
static frame (in our case, the velocity at which the PBH moves away
from the Earth), the redshift factor $\alpha$ can be rewritten as
\begin{equation}\label{red2}
\alpha = \sqrt{\frac{f(R)}{f(r_0)}}
\left(1-\beta \cos\theta_{\rm e}^{'}\right).
\end{equation}
After some simple algebra, we obtain for the first
term on the right hand side
of equation~(\ref{red2})
\begin{eqnarray}\label{red3}
\sqrt{\frac{f(R)}{f(r_{\rm o})}} &=&
\sqrt{\frac{f(r_{\rm e})}{f(r_{\rm o})}}
\sqrt{\frac{f(R)}{f(r_{\rm e})}}
= \sqrt{\frac{f(r_{\rm e})}{f(r_{\rm o})}}
\left[1-\left(1-\frac{f(r_{\rm e})}{f(R)}\right)\right]^{-1/2}
\nonumber\\
&=& \sqrt{\frac{f(r_{\rm e})}{f(r_{\rm o})}}
\left(1-\beta^2\right)^{-1/2},
\end{eqnarray}
where the first factor in equation~(\ref{red3})
represents the gravitational redshift between the emitter
and the observer $\gamma_{\rm G}$, and the second one
is the kinematic factor $\gamma = \left(1-\beta^2\right)^{-1/2}$.
So, the transformation relation can expressed as
\begin{equation}\label{PBHtrans}
T_{\rm P}^{({\rm G})} = \alpha T_{\rm P}^{'} =
\gamma_{\rm G} \gamma \left(1-\beta
\cos\theta_{\rm e}^{'}\right) T_{\rm P}^{'}.
\end{equation}

Please note that the latter expression for
$T_{\rm P}^{({\rm G})}$
is equivalent to equation~(16)
in the original paper \citep{Barco2021},
with the exception of the emission angle
$\theta_{\rm e}^{'}$ (in the corrected version,
no critical angle $\theta_{\rm c}^{'}$ is considered).
Identical result can be easily deduced from
equation~(14) of \cite{McMaken2022}, simply by considering
the emission angle transformation relation
(equation (41) of \cite{Yoshino2019})
\begin{equation}\label{emtrs}
\cos\theta_{\rm e} = \frac{\cos\theta_{\rm e}^{'} - \beta}
{1- \beta \cos\theta_{\rm e}^{'}},
\end{equation}
and performing some elementary calculations.

\begin{figure}
\begin{center}
\includegraphics[width=.49\textwidth]{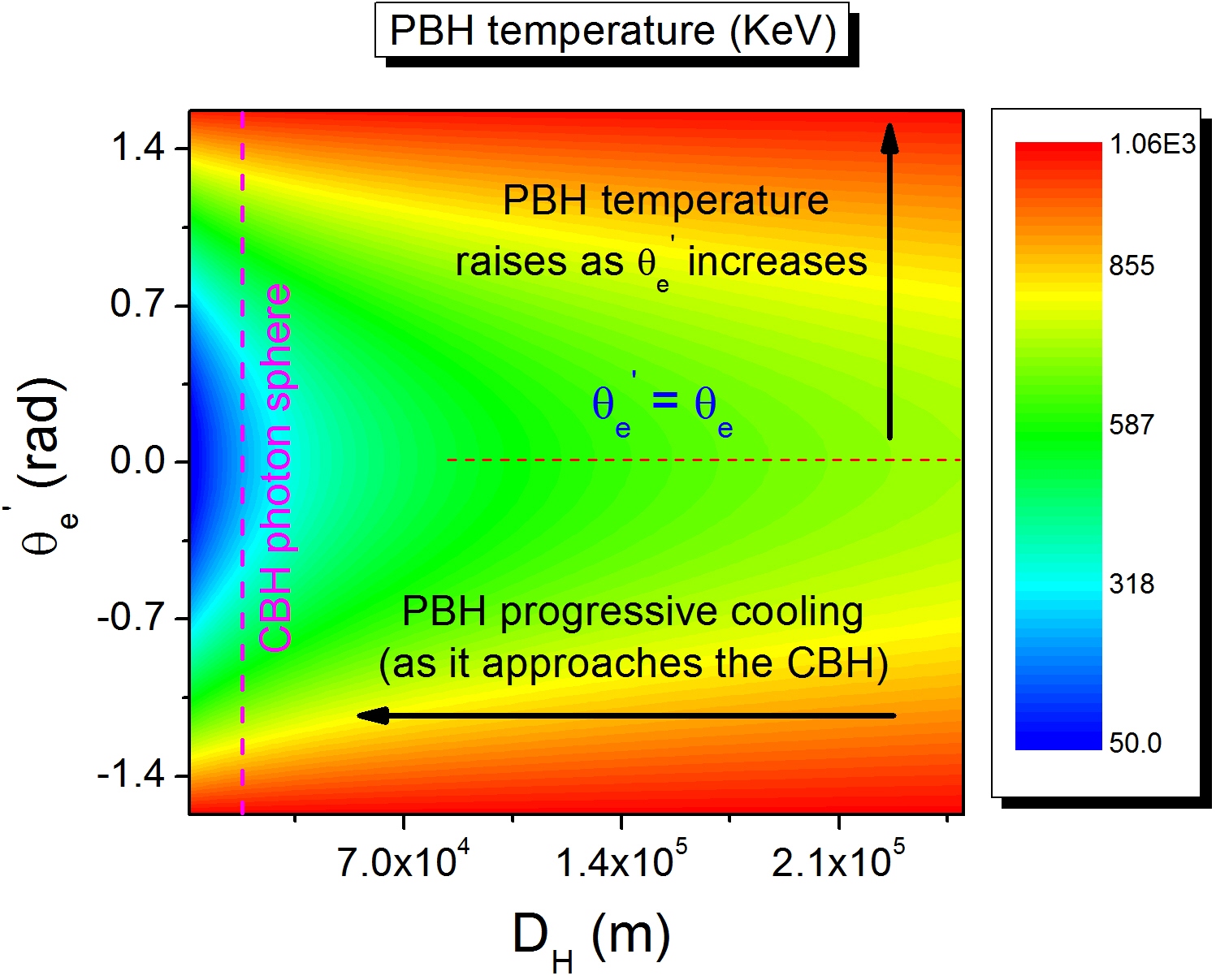}
\caption{Hawking temperature $T_{\rm P}^{({\rm G})}$
of an asteroid-mass PBH
($M_{\rm P} \simeq 5 \times 10^{-18} M_{\sun}$)
as a function of its emission angle $\theta_{\rm e}^{'}$
and the binary BH horizon separation $D_{\rm H}$,
calculated via equation~(\ref{PBHtrans}).
The light black hole describes a free fall from
rest at infinity ($E_{\rm s}=1$ and $L_{\rm s}= 3.3
\times 10^{-4}$ m, in geometrized units)
towards a 10 $M_{\sun}$ central black hole.
The horizontal dashed line (where $\theta_{\rm e}^{'} =
\theta_{\rm e} = 0$ rad) depicts the Earth's line of sight
(please, see again Fig.~\ref{fig1}).}
\label{fig2}
\end{center}
\end{figure}

After implementing the corrections, a possible astrophysical
scenario includes an asteroid-mass PBH \citep{Wang2021, Coogan2021}
with $M_{\rm P} \simeq 5 \times 10^{-18} M_{\sun}$
describing a free fall from rest at infinity
towards a 10 $M_{\sun}$ CBH at 48.6 kpc.
In Fig.~\ref{fig2}, the vertical axis now represents
the emission angle $\theta_{\rm e}^{'}$
(provided that, in the current situation,
the PBH specific energy $E_{\rm s}=1$ with
specific angular momentum $L_{\rm s}= 3.3 \times 10^{-4}$ m,
both in geometrized units).

It can be observed that,
for a given emission angle $\theta_{\rm e}^{'}$,
the PBH Hawking temperature diminishes as the light
BH approaches its heavier companion.
This cooling behaviour can be easily understood after inspection of
equation~(\ref{PBHtrans}): as the PBH gains velocity during its free fall,
the term $1-\beta \cos\theta_{\rm e}^{'}$
tends progressively to zero when the emission angle $\theta_{\rm e}^{'} =
\theta_{\rm e} = 0$ rad (i.e., along the Earth's line of sight).
As a numerical example, when the asteroid-mass PBH is located at
$10^{5}$ m above the CBH horizon, $\beta = 0.48$c, $\gamma = 1.14$
and $\gamma_{\rm G}=0.88$ with an observed Hawking temperature of
$552.5$ keV. At $10^{3}$ m above the horizon,
$\beta = 0.98$c, $\gamma = 5.53$ and $\gamma_{\rm G}=0.18$, where now
$T_{\rm P}^{({\rm G})} = 17.5$ keV. It should be underlined that the
emitted PBH Hawking radiation is not Lorentz-boosted
(as incorrectly commented in the original version).

\begin{figure}
\begin{center}
\includegraphics[width=.49\textwidth]{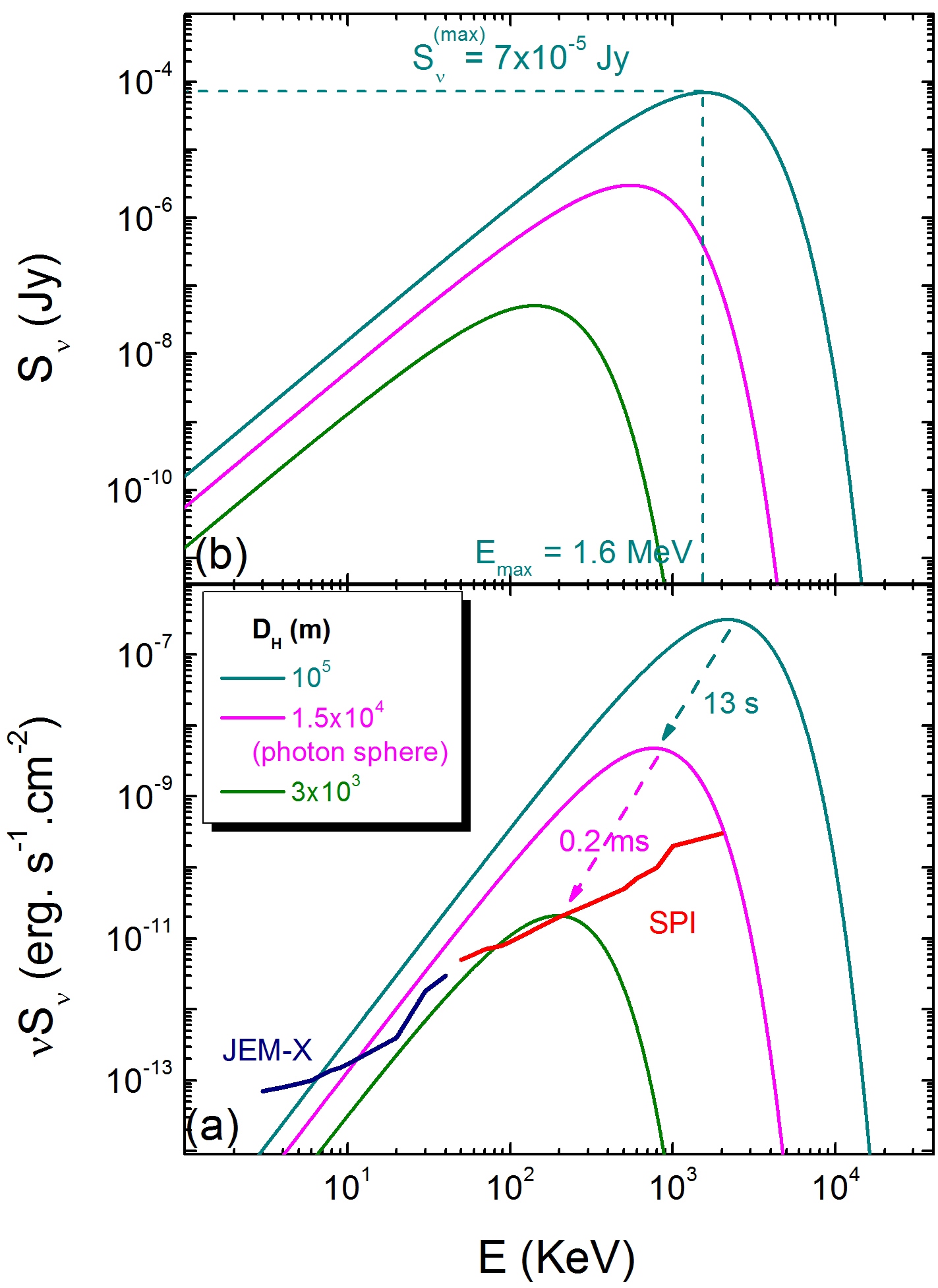}
\caption{(a) Fluence spectrum $\nu S_{\nu}$ for different
horizon separations $D_{\rm H}$
of an asteroid-mass PBH
($M_{\rm P} \simeq 5 \times 10^{-18} M_{\sun}$)
describing a free fall towards a 10 $M_{\sun}$
central black hole at 48.6 kpc, along the Earth's
line of sight (numerical results performed via
equation~(\ref{nuSnu})).
As an indicative manner,
the sensitivities of the Joint European X-Ray
Monitor (JEM-X) and INTEGRAL spectrometer (SPI)
for 1 year are illustrated, (b) the corresponding
PBH flux density values $S_{\nu}$ obtained numerically from
equation~(\ref{Rnu}) (taking into account
the solid angle $\Omega$ subtended by our PBH).}
\label{fig3}
\end{center}
\end{figure}

The numerical study of the PBH fluence spectrum
$\nu S_{\nu}$ (via equation~(\ref{nuSnu}) of this revised version)
and the associated flux density $S_{\nu}=\Omega R_{\nu}$,
for different horizon separations $D_{\rm H}$ is represented
in Fig.~\ref{fig3}.
These parameters strongly depend on the solid angle $\Omega$
subtended by the PBH (considered as a point source) and obey
the well-known inverse square law.
So, it is assumed that the beginning of the GRB occurs when
the fluence spectrum $\nu S_{\nu}$ is sufficiently high
to be detected by ground-based or space observatories
(that is, when the PBH is close enough to Earth).
In our updated model, this happens when $D_{\rm H}= 10^{5}$ m
(please, see again Fig.~\ref{fig3}(a)) while the final stage
of the burst occurs at about 3 km above the CBH horizon,
when $\nu S_{\nu}$ drops below the sensitivity of the
Joint European X-Ray Monitor (JEM-X).

The astrophysical scenario is the same as
the previously described in Fig.~\ref{fig2},
with an emission angle $\theta_{\rm e}^{'} =
\theta_{\rm e} = 0$.
A decreasing $\nu S_{\nu}$ fluence spectrum
is also observed as the horizon separation
$D_{\rm H}$ decreases, with measurable time
intervals $\Delta t$ between different PBH
approximations. Unlike the original version, the
closest approach to the CBH event horizon
contributing to the observable GRB is now about 3 km.
It should be also noted that the coordinate time interval
is clearly reduced as the PBH gains speed during its free fall,
achieving a velocity of 0.82c at the CBH photon sphere.

\begin{figure}
\begin{center}
\includegraphics[width=.48\textwidth]{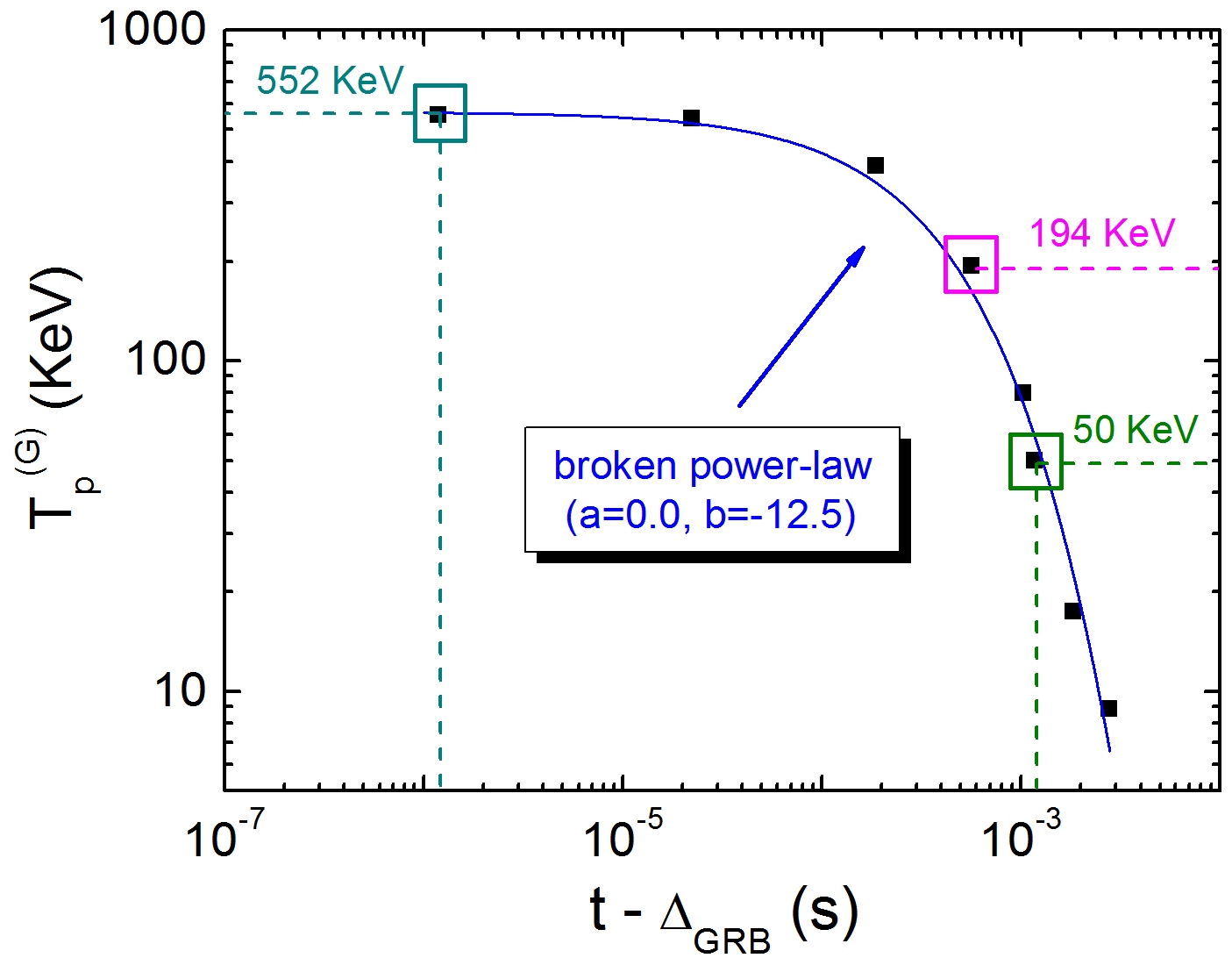}
\caption{The cooling behaviour of the PBH Hawking
temperature $T_{\rm P}^{({\rm G})}$
as a function of the coordinate time $t$
(once the estimated GRB duration $\Delta_{\rm GRB} = 13.2$ s
is subtracted). The PBH Hawking temperatures for the three
horizon separations in Fig.~\ref{fig3}
are also shown. The solid line stands for
the broken power-law with the selected parameters
$a=0.0$, $b=-12.5$, $t_0=9 \times 10^{-3}$ s,
$\delta=1.0$ s, $t_n=1.5 \times 10^{-3}$ s and
$T_n = 36$ keV.}
\label{fig4}
\end{center}
\end{figure}

In connection with the numerical results
depicted in Fig.~\ref{fig3}, the cooling behaviour
of the PBH Hawking temperature $T_{\rm P}^{({\rm G})}$
is described in Fig.~\ref{fig4}. For a better visualization,
the horizontal axis now represents the coordinate time $t$
once the estimated GRB duration ($\Delta_{\rm GRB} = 13.2$ s)
is subtracted. In this revised version, the broken
power-law parameters correspond to $a=0.0$, $b=-12.5$,
$t_0=9 \times 10^{-3}$ s, $\delta=1.0$ s,
$t_n=1.5 \times 10^{-3}$ s and $T_n = 36$ keV.
It is also worth mentioning that the slope index $b$
(which takes into account the cooling process at the
final stage of the GRB) is steeper than in the
original version. This is due to the above-mentioned
shortening of the time intervals as the PBH is
more accelerated.

In relation to the capture rate of a PBH by
a more massive BH, \cite{McMaken2022} recently
claimed an implausible occurrence based on
\cite{Kouvaris2008} and \cite{Capela2013} research.
Let us analyze in detail if McMaken's arguments
and conclusions (concerning such capture rate)
can be properly applied to the binary black hole
model here presented.

According to \cite{McMaken2022}, the capture rate $\mathcal{F}$
is estimated by integrating a Maxwellian dark matter
distribution, with velocity dispersion $\bar{v}$,
over the $E_{\rm s}$ and $L_{\rm s}$ space
with well-defined integration limits (in accordance
with equation~(3) of \cite{Kouvaris2008}, within the context
of weakly interacting massive particles (WIMPs)
captured by a compact star).
It must be borne in mind that, within Kouvaris and Capella's
formalism, the capture rate $\mathcal{F}$ is derived
for a WIMP (or PBH, in Capela's paper) which loses its initial
energy due to the accretion of star's material and dynamical
friction, so it becomes gravitationally bound
(please, see Section IIa of \cite{Capela2013}).

Consequently, as stated by \cite{Kouvaris2008},
the capture rate $\mathcal{F}$ should be calculated
in two steps: first, it must be determined
what part of the $E_{\rm s}$ and $L_{\rm s}$ space
can give orbits for the WIMps that intersect with the
neutron star (this is what McMaken is supposed to calculate).
A second step involves a derivation of the fraction of
such particles that lose enough energy to be trapped
inside the star. 

In that sense, equation~(16) of \cite{McMaken2022} (the alleged
expression for calculating the capture rate of a PBH by a heavier
black hole) is exactly the same as equation~(13) of \cite{Capela2013}
in the context of a neutron star of radius $R$ (which, in turn, is also
equivalent to equation~(11) of \cite{Kouvaris2008}, once the
integration of the Maxwellian distribution is carried out).
Please note that Capela's equation comes from a particular regime,
where the PBH energy loss $E_{\rm loss}$ is particularly high
(i.e., when $E_{\rm loss} \gg m_{\rm PBH} \bar{v}^{2}/3$,
as well explained by \cite{Capela2013} in Section IIb). 
How can Capella's approach be valid for our binary BH scenario?

Even assuming that equation~(16) of \cite{McMaken2022}
could properly be applied to our model, let us
recalculate the supposed PBH capture rate $\mathcal{F}$
with the same parameters as McMaken's paper. In this reference,
all dark matter is composed of PBHs (that is,
$\Omega_{\rm PBH} = \Omega_{\rm DM}$) with a current observed
DM density of $\rho_{\rm DM} = 0.5$ $\mathrm{GeV.cm^{-3}} =
8 \times 10^{-5} \mathrm{J.m^{-3}}$. The velocity dispersion
parameter is $\bar{v}=7000$ $\mathrm{m.s^{-1}}$
and a selected radius of capture $R = 1$ kpc from Earth.
Considering an all-DM primordial black hole
of $5 \times 10^{17}$ kg
\citep{Inomata2017, Bartolo2019, deLuca2021}
and a 10 $M_{\sun}$ central black hole with an Schwarschild radius
of $2.9 \times 10^{4}$ m (this last parameter can be easily
replaced by other massive CBHs with identical consequences),
the resulting capture rate is $\mathcal{F} = 8.6 \times 10^{-2} \mathrm{s}
= 2.7 \times 10^{6} \mathrm{yr^{-1}}$ (according to equation~(16)
of \cite{McMaken2022}). This result is 15 orders of magnitude
greater than the reported by McMaken, proving explicitly the inconsistency
of such capture rate calculations.

As a final remark, it is worth noting that the astrophysical
scenario described by \cite{Barco2021} does not intend to reinterpret
the thermal GRBs described satisfactorily by the fireball model
\citep{Ghirlanda2003}, but provide an alternative explanation of
such astrophysical events based on PBH origin.

\section*{Acknowledgements}

The author gratefully acknowledges the anonymous reviewer
for helpful guidance throughout the refereing process.

O. del Barco also thanks T. McMaken (despite some unfounded
criticism commented in this \emph{Erratum}) for pointing out
some mistakes of my astrophysical model (properly
corrected in this updated version), on the basis of
constructive and collegial scientific debate.

\bibliographystyle{mnras}

\bsp
\label{lastpage}
\end{document}